\documentclass[
aip,
amsmath, amssymb,
reprint,
]{revtex4-2}
\usepackage{graphicx}
\usepackage{dcolumn}
\usepackage{bm}
\usepackage[utf8]{inputenc}
\usepackage[T1]{fontenc}
\usepackage{mathptmx}
\usepackage{etoolbox}
\newcommand{\vect}[1]{\boldsymbol{#1}} 

\usepackage{color}

\setcitestyle{notesep={; },round,aysep={},yysep={;}}

\begin{document}
	
	\preprint{AIP/123-QED}
	
	\title{A machine learning model to classify dynamic processes in liquid water}
	\author{Jie Huang}
	\affiliation{Jie Huang, Prof. Dr. Shiben Li\\ \ \ \ Department of Physics, Wenzhou University, Wenzhou, Zhejiang 325035, China; E-mail: shibenli@wzu.edu.cn}
	
	\author{Gang Huang*}
	\affiliation{Dr. Gang Huang\\ \ \ \ Institute of Theoretical Physics, Chinese Academy of Sciences, Beijing 100190, China; E-mail: hg08@lzu.edu.cn
	}

	\author{Shiben Li*}
	\affiliation{Jie Huang, Prof. Dr. Shiben Li\\ \ \ \ Department of Physics, Wenzhou University, Wenzhou, Zhejiang 325035, China; E-mail: shibenli@wzu.edu.cn}
	
	\begin{abstract}
	The dynamics of water molecules plays a vital role in understanding water. We combined computer simulation and deep learning to study the dynamics of H-bonds between water molecules. Based on \emph{ab initio} molecular dynamics simulations and a newly defined directed Hydrogen (H-) bond population operator, we studied a typical dynamic process in bulk water: interchange, in which the H-bond donor reverses roles with the acceptor. By designing a recurrent neural network-based model, we have successfully classified the interchange and breakage processes in water. We have found that the ratio between them is approximately 1:4, and it hardly depends on temperatures from 280 to 360 K. This work implies that deep learning has the great potential to help distinguish complex dynamic processes containing H-bonds in other systems.
	\end{abstract}

	\maketitle
	
	\section{Introduction}
	As one of the big questions in the 21st century \cite{Kennedy2005}, the structure of water is essential for understanding cells, biological processes, and ecosystems \cite{Franks2000, Pal2004, Chaplin2006, Ball2017}. Water's surprising properties \cite{Stillinger1980, Errington2001, Dum2020}, such as increased density on melting, high surface tension, maximum density at 4 $^\circ$C, are closely related to the H-bonds\cite{Kumar2007, Nilsson2015, Mundzeck2020}. Despite the fact that it is tough to capture the ultrafast motion of atoms during dynamic processes \cite{Karamatskos2019}, watching water molecules as they dance is the key to understand the dynamic properties of water \cite{Perakis2018} from the molecular level.  Many methods over the past three decades were used to study water molecules' motion, such as scanning tunneling microscopy (STM) \cite{Mitsui2002,Kumagai2008}, femtosecond pump-probe \cite{Woutersen1997}, infrared (IR) spectroscopy \cite{Bakker2002, Fecko2003, Karamatskos2019, Inoue2020}, X-rays \cite{Iwashita2017, Perakis2018, Loh2020}, neutron scattering \cite{HeadGordon2002},  and computer simulations \cite{Ranea2004, Laage2006, Kumar2007,  Fang2020}. 

In this work, we focus on one specific dynamic process \textit{in bulk water}: interchange \cite{Keutsch2001}, in which the H-bond donor reverses roles with the acceptor in the same H-bond. This process was observed in the gas-phase water dimer by Saykally and coworkers. The interchange process, which involves the quantum tunneling effect \cite{Fellers1999, Keutsch2001, Ranea2004, Kumagai2008, Fang2020}, is essential for understanding water molecules' dynamics. Also, since interchange processes are closely related to the H-bond network dynamics, it is likely to play a critical role in biological processes, like proton transfer
\cite{Agmon1995, Thomaston2017,Gelenter2021}. So far, interchange processes have been found in water dimer adsorbed on metal surfaces \cite{Kumagai2008}. Using \emph{ab initio} molecular dynamics (AIMD) simulations \cite{Kuhne2020}, Ranea et al. \cite{Ranea2004} found that the interchange process can be used to explain the rapid diffusion behavior of water dimer on the Pd(111) surface. Fang et al. \cite{Fang2020} found that interchange process is a mechanism of the rapid movement of water dimers on metal surfaces. As for bulk water, Lagge and Hynes found that the redirection of water molecules involves large-angle jumps \cite{Laage2006}, which involves the redirection of \textit{multiple} water molecules referred to as \textit{H-bond exchange}, and it is supported by the subsequent experiments \cite{Moilanen2009, Ji2010}. 

The interchange process involves the concerted rotation of both water molecules engaged in a H-bonded pair. This mechanism is important in small clusters where the future hydrogen-bond donor OH group is typically initially dangling. There are some simulation studies on the interchange process in water clusters \cite{Fellers1999, Keutsch2001, Schulz2018, Samala2019, Mndez2020}, as far as we know, the question of the ratio of interchange to other dynamic processes related to H-bonds \textit{in bulk water} has not been discussed. To determine the proportion of interchange processes, we simulated bulk water in a canonical (NVT) ensemble using a specific AIMD simulation method: the density functional molecular dynamics (DFTMD) simulation. We observed interchange processes in bulk water by analyzing the dynamic trajectory. 

As it's tough to quantify interchange processes in a large number of ultrafast dynamic processes in liquid water, we have designed a recurrent neural network (RNN)-based model to classify the H-bond dynamic processes. Unlike general classification methods, this model has the capability of classifying the \emph{dynamic processes} related to H-bonds in bulk water. Using this model, we have obtained the relative ratio of interchange and breakage processes in bulk water and explored the effect of temperature on this ratio. Our work presents the great capacity to use the RNN-based deep learning method to study the dynamic properties of liquid water.

The aims of this work is to provide a machine learning-based model to classify dynamic processes and to determine the proportion of interchange processes in water as one useage of the model. The organization of the paper is the following. We present the results and discussion in Sec.\thinspace\ref{result}. At first, the dynamic graph representation of H-bond networks is introduced in  \ref{ssec:dgraph} and the main characteristics of interchange processes are obtained in \ref{ssec:da_exchange}. Then we implement the RNN-based classifier for different types of dynamic processes in liquid water in \ref{ssec:rnn} and explore the temperature dependence of the relative ratios of interchange and breakage processes in \ref{ssec:proportions};   The discussion of two factors, the mean number of H-bonds and the rate of breakage and formation of H-bonds in liquid water, related closely to the temperature dependence are discussed in \ref{ssec:trend}.  Finally, we present the methods details  and  conclusions of our study in Sec.\thinspace\ref{methods} and \thinspace\ref{conclusions}, respectively.

\section{Results and discussion}\label{result}
\subsection{Dynamic graph representation of H-bond networks} \label{ssec:dgraph}
\begin{figure}[tbhp]
	\centering
	\includegraphics[width=\linewidth]{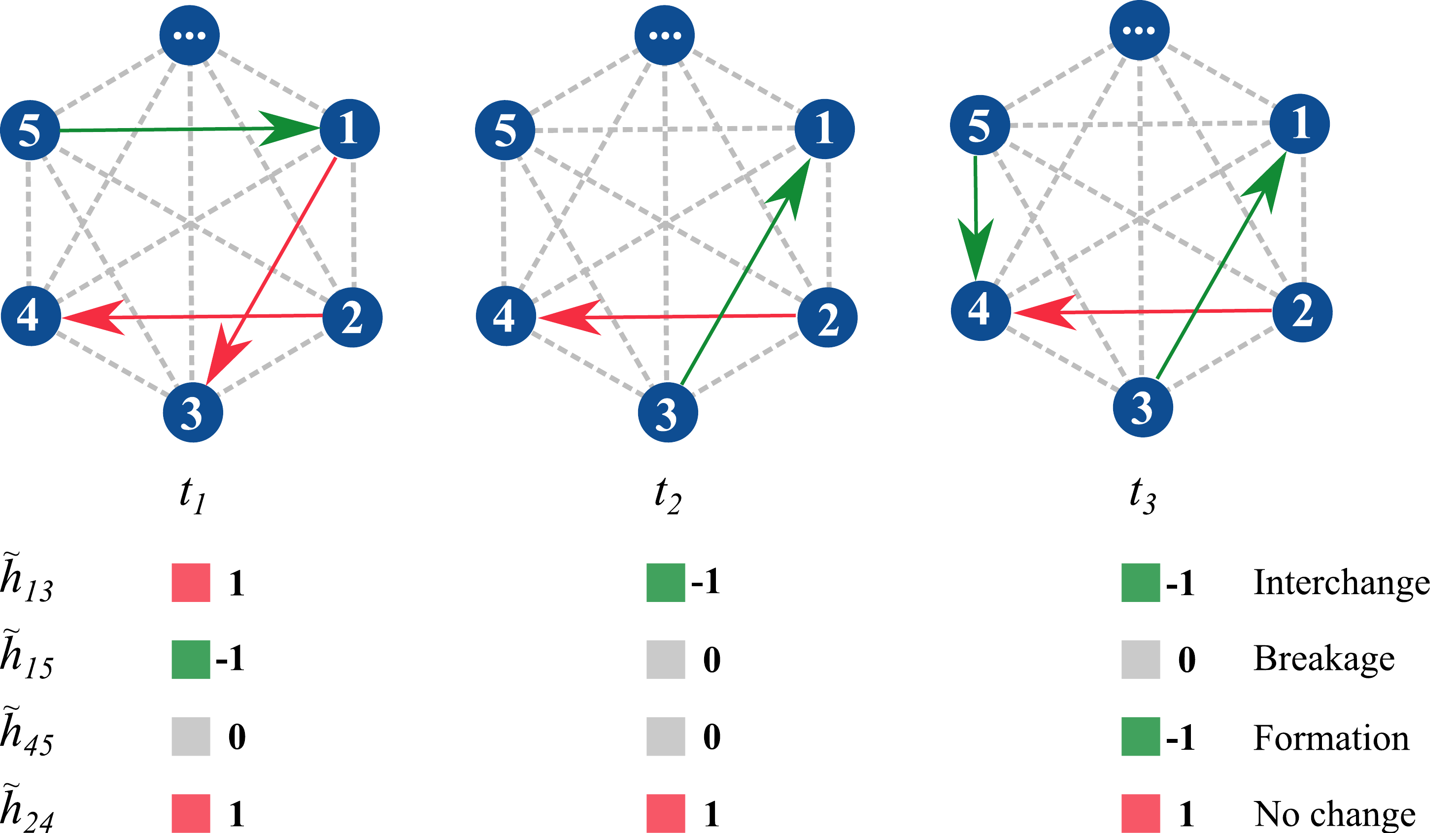}
	\caption{Dynamic graph representation of the H-bond network in simulated bulk water. Nodes represent water molecules; solid red or green arrows represent H-bonds; and dashed grey lines represent Q-bonds.  The colors red, grey, and green indicate $\tilde{h}_{ij}$=$1$, $\tilde{h}_{ij}$=$0$, and $\tilde{h}_{ij}$=$-1$, respectively. From the time sequence of $\tilde{h}_{ij}$, we know how the H-bond configuration of $b_{ij}$ changes over time. Four typical H-bond configuration change processes are illustrated for $b_{13}$, $b_{15}$,  $b_{45}$,  and $b_{24}$, corresponding to interchange, breakage, formation, and no change, respectively.}
	\label{fig:dynamic_graph}
\end{figure}

As shown in Fig.\thinspace\ref{fig:dynamic_graph}, a directed \emph{dynamic graph} is used to describe the bulk water system of $N$ water molecules. Each water molecule may form an H-bond with any of the remaining $N$-$1$  molecules. For convenience, we call any pair of water molecules $(i, j)$ a \emph{quasi}-hydrogen bond (Q-bond), denoted as $b_{ij}$ and represented as a dashed line in Fig.\thinspace\ref{fig:dynamic_graph}. 

Inspired by Luzar and Chandler's H-bond population operator \cite{Luzar1996}, we define a \emph{directed} H-bond population operator $\tilde h_{ij}$ for $b_{ij} \ (i < j)$ at time $t$ as Eq.\thinspace\ref{eqn:h_ij}.  

\begin{align}
	\tilde h_{ij}(t)=\left\{
	\begin{array}{rcl}
		1       &      & {\text{H-bonded, $i$ is the donor}}\\
		0    &      & {\text{Not H-bonded}} \\   \label{eqn:h_ij}
		-1       &      & {\text{H-bonded, $j$ is the donor}}
	\end{array} \right.
\end{align}

We know from $\tilde h_{ij}$ whether an H-bond exists in  $b_{ij}$ and the donor-acceptor pair of the formed H-bond. At the bottom of Fig.\thinspace\ref{fig:dynamic_graph}, we demonstrate four typical H-bond configuration change processes by using the sequences of $\tilde h$: interchange, breakage, formation, and no change. Besides, the Q-bonds likely to form H-bonds are the most relevant water molecule pairs to the breakage and reforming of H-bond networks. The following geometric criteria  \cite{Sciortino1989, Balasubramanian2002, Michaud-Agrawal2011} of an H-bond is used:  O-O distance $R_{\text{OO}^{\prime}} < R_\text{cutoff} = 3.5 \text{ \AA}$ and angle $\text{O-H}\cdots \text{O} > \theta_\text{cutoff} = 120^{\circ}$.
\begin{figure}[h]
	\centering
	\includegraphics[width=\linewidth]{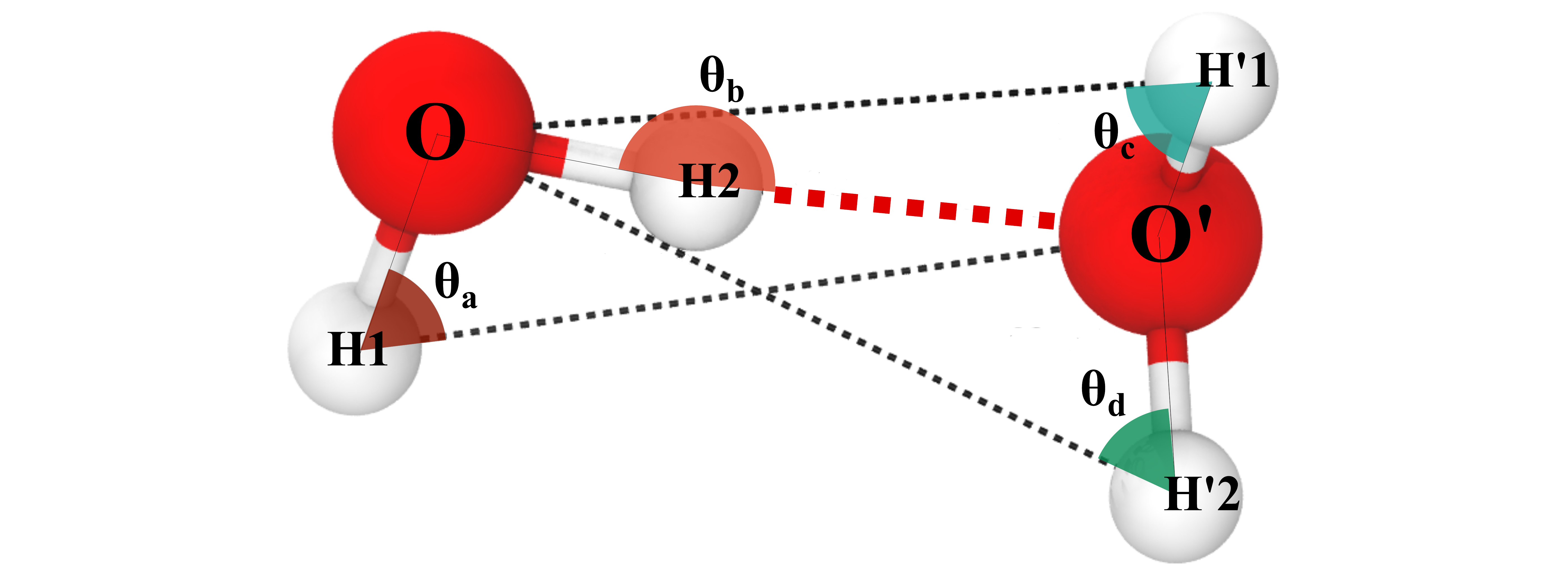}
	\caption{
		Scheme of the geometric coordinates. $R_{\text{OO}'}$ is the O-O distance. Four angles $\widehat{\mathrm{O} \mathrm{H1O^{\prime}}}$, $\widehat{\mathrm{O} \mathrm{H2O^{\prime}}}$ , $\widehat{\mathrm{O^{\prime}} \mathrm{H^{\prime}1O}}$, and $\widehat{\mathrm{O^{\prime}} \mathrm{H^{\prime}2O}}$ are represented as  $\theta_\text{a}$, $\theta_\text{b}$, $\theta_\text{c}$, and $\theta_\text{d}$, respectively.
		If $R_{\text{OO}'} < 3.5 \text{ \AA}$ , and any angle $\theta > 120^{\circ}$ ($\theta \in \{\theta_\text{a}$,$\theta_\text{b}$, $\theta_\text{c}$, $\theta_\text{d}\}$), then an H-bond exists in this Q-bond.  Here, the oxygen atom O  as a donor donates the hydrogen atom H2 to the acceptor $\text{O}^{\prime}$. Since $R_{\text{OO}^{\prime}}< 3.5 \text{ \AA}$  and $\theta_\text{b}> 120^{\circ}$, we describe this state of $b_{\text{OO}^\prime}$ at this time $t$ by $\tilde h_{\text{OO}^{\prime}}(t) = 1$.
	}\label{fig:mesurements}
\end{figure}
As shown in Fig.\thinspace\ref{fig:mesurements}, $R_{\text{OO}^{\prime}}$, $\theta_\text{a}$, $\theta_\text{b}$, $\theta_\text{c}$, and $\theta_\text{d}$ are monitored for Q-bonds to study the reorientation and breakage mechanism of H-bonds. 

\subsection{Interchange process}\label{ssec:da_exchange}
\begin{figure*}
	\centering
	\includegraphics[width=\linewidth]{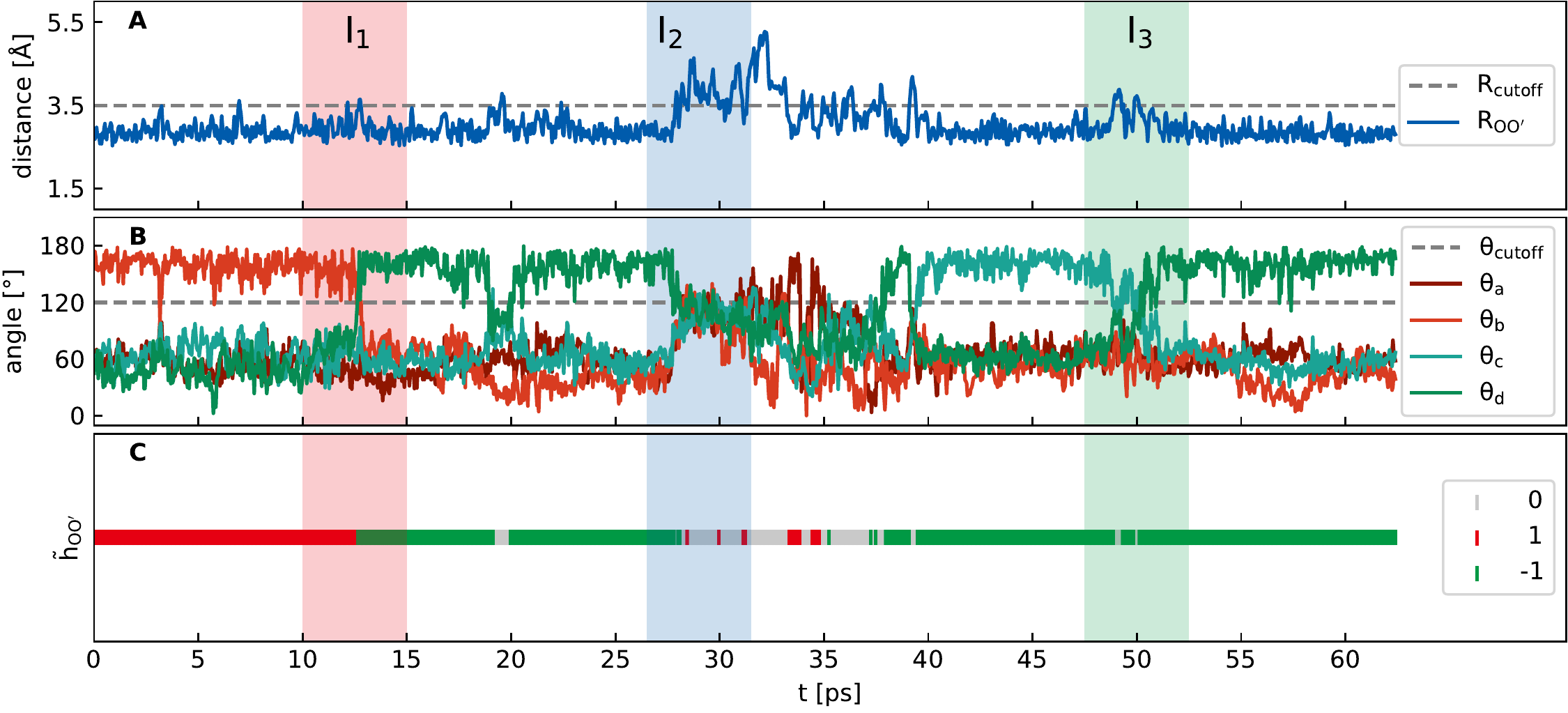}
	\caption{
		Interchange ($\text{I}_1$), breakage ($\text{I}_2$), and bifurcation rearrangement ($\text{I}_3$) process for one typical Q-bond in bulk water. When an H-bond exists in a Q-bond if $\theta_\text{a} > \theta_\text{cutoff}$ or $\theta_\text{b} > \theta_\text{cutoff}$, then the oxygen atom O is the donor; else, if $\theta_\text{c} > \theta_\text{cutoff}$ or $\theta_\text{d} > \theta_\text{cutoff}$, then the oxygen atom O$^\prime$ is the donor. Three typical processes are interchange, where the water molecule pairs exchange their roles as H-bond donor and acceptor; breakage, where the H-bond is breaking as the distance increase of  this water molecule pair; and  bifurcation rearrangement, where the donated hydrogen atom of the H-bond donor exchanged. Through $\tilde{h}$, we can see whether an H-bond exists between a Q-bond, also know the donor and acceptor if an H-bond exists. In panel (C), the grey, red, and green lines indicate the $\tilde{h}_{{\text{OO}}^\prime}$ states.}\label{fig:da_dha}
\end{figure*}
The AIMD simulation trajectory allows us to observe the details of the H-bond dynamics. Figure \ref{fig:da_dha} demonstrates the dynamics of the distance, angles, and directed H-bond population for $b_{\text{OO}^\prime}$. 
Intervals $\text{I}_1$, $\text{I}_2$,  and $\text{I}_3$ correspond to three typical H-bond dynamic processes.  \textit{interchange} ($\text{I}_1$): We notice $\theta_\text{b}> \theta_\text{cutoff}$ in the first half and $\theta_\text{d}> \theta_\text{cutoff}$ in the second half. Besides,  $\tilde h_{\text{OO}^{\prime}}$ changes from $1$ to $-1$, indicating that the donor and acceptor have exchanged. 
\emph{breakage} ($\text{I}_2$):  $\tilde h_{\text{O}\text{O}^{\prime}}=-1$ in the first half of  $\text{I}_2$,  and $\tilde h_{\text{O}\text{O}^{\prime}}=0$ for most of the second half. There is no H-bond in the second half because $R_{\text{OO}^{\prime}}>R_\text{cutoff}$, i.e., the increase of distance $R_{\text{OO}^{\prime}}$ causes the H-bond to break. 
\emph{bifurcation rearrangement} motion ($\text{I}_3$)\cite{Keutsch2001, Brown1998}: At first, the hydrogen atom $\text{H}^{\prime}1$ is donated to form an H-bond as $\theta_\text{c}> \theta_\text{cutoff}$. Then $\theta_c$ decreases and $\theta_d$ increases until $\theta_\text{d}> \theta_\text{cutoff}$,  i.e.,  the other hydrogen atom $\text{H}^{\prime}2$ of the donor is donated to form the H-bond. Therefore, the hydrogen atom contributed by the donor is changed. Because of the identity of hydrogen atoms, it is impossible to distinguish the configuration of water molecules before and after the process. However, during the interchange process, the direction of the water molecules' dipole moment will change, indicating that the water molecules' microscopic configuration will change. So in the rest of the article, we focus on the interchange process. 

\begin{figure}[thb]
	\centering
	\includegraphics[width=\linewidth]{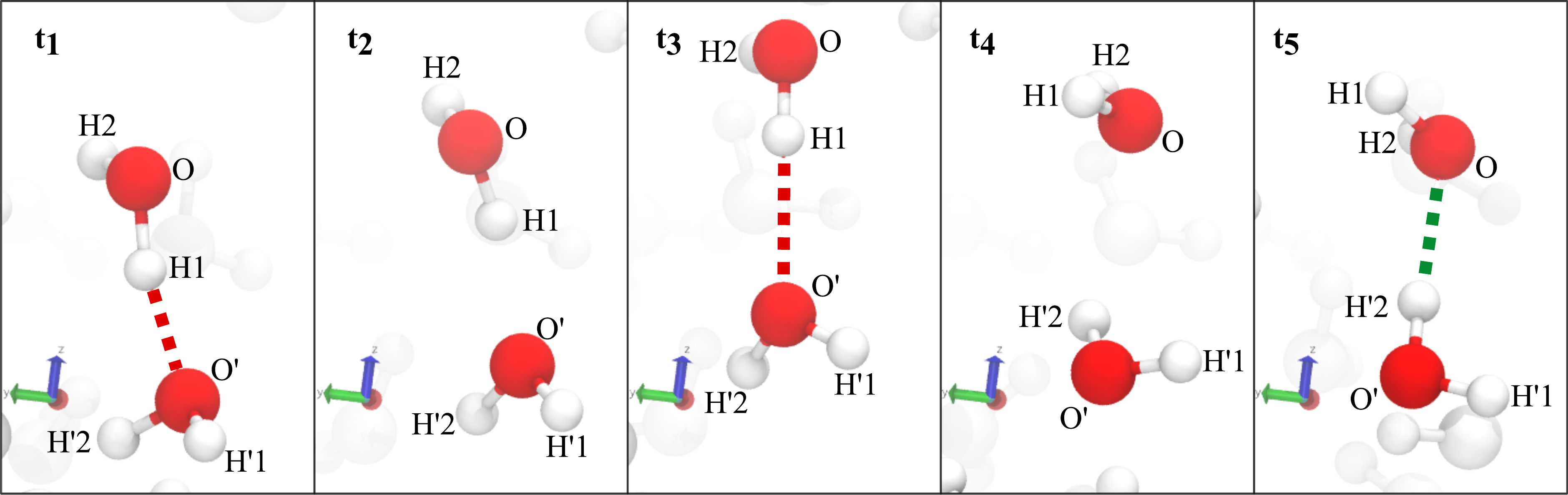}
	\caption{A typical interchange process, where two water molecules exchange their roles as H-bond donor and acceptor via water molecules' reorientation in an concerted manner. The donor oxygen atom has changed from the original O to O$^\prime$ (color of dashed line changed from red to green). Besides, we have also noticed that the H-bond briefly breaks during the interchange process, causing the fluctuation of the $\tilde{h}$ sequence.}\label{fig:da_demo}
\end{figure}
Figure \ref{fig:da_demo} shows a typical interchange process in water (see Fig.\thinspace$\text{S4}$ in SI Sec.\thinspace3 and movies in supplementary material for more H-bond configuration change processes). A dashed line represents an H-bond, and its color (red or green) indicates its direction. Using $\tilde h$, we can describe the H-bond configuration change progress without paying attention to the distance and angles. Therefore, $\tilde h$ dramatically simplifies the description for the H-bond configuration change process. Nevertheless, during dynamic processes, the fluctuations of $\tilde h$ that can result from the vibration of water molecules will bring a huge challenge for the classification of H-bond configuration change processes. In addition, due to a large number of Q-bonds in the simulated bulk water, finding a specific H-bond configuration change process in 60 ps is like finding a needle in a haystack. Therefore, we design an RNN-based model that recognizes the dynamic processes related to H-bonds and uses it to determine various processes in water, thereby determining the ratio of interchanges.

\subsection{RNN-based classifier for H-bond configuration change process}\label{ssec:rnn}
\begin{figure*}[t]
	\centering
	\includegraphics[width=1\linewidth]{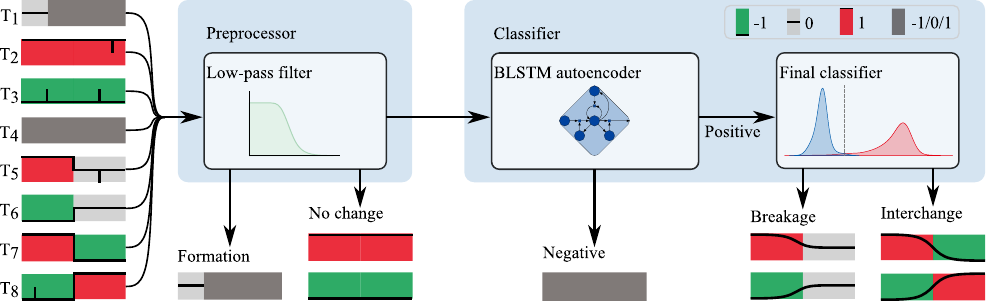}
	\caption{The processing flow of the H-bond configuration change classifier based on RNN. (i). Different types of $\tilde{h}$ sequences: $\text{T}_1$: Formation or no H-bond; $\text{T}_2$, $\text{T}_3$: No change; $\text{T}_4$: Negative sequence;  $\text{T}_5$, $\text{T}_6$: Diffusion;  $\text{T}_7$, $\text{T}_8$: interchange. We refer to the sequences of breakage and interchange as positive sequences. (ii). The preprocessor filters out the high-frequency components of $\tilde{h}$ and excludes $\text{T}_1$, $\text{T}_2$, and $\text{T}_3$.  (iii). The classifier consists of a BLSTM AE to separate the positive and negative sequences and a final classifier to distinguish breakage and interchange sequences.}
	\label{fig:flow_chart}
\end{figure*}

We can see the interchange and breakage processes intuitively from $\tilde{h}$. Specifically, in the interchange process, $\tilde{h}$ changes from $\pm 1$ to $\mp 1$; in the breakage process, $\tilde{h}$ changes from $\pm 1$ to 0. Therefore, in principle, by observing the sequence of $\tilde{h}$ within a time window, we can classify the H-bond configuration changes during this period. Although we can see some change patterns in interchange and breakage processes, it is still challenging to distinguish different $\tilde{h}$ sequences due to the fluctuation. Therefore, we have designed a processing flow to classify the H-bond configuration change process based on RNN, as shown in Fig.\thinspace\ref{fig:flow_chart}.
In the preprocessor, we use a low-pass filter to filter out the high-frequency fluctuations of $\tilde{h}$ sequences. As we focus on the configuration change processes of H-bonds,  we exclude sequences without H-bonds at the beginning ($\text{T}_1$) and the sequences whose H-bond configuration are unchanged ($\text{ T}_2$, $\text{T}_3$) according to the initial value and the variance of $\tilde{h}$ sequences (see Methods section).
After preprocessing, the task we need to deal with is a time series classification problem: In addition to interchange and breakage processes, there are also many irregular and complicated processes. We call the sequences of interchange and breakage \textit{positive} and all sequences other than these two types \textit{negative} for convenience. Negative sequences ($\text{T}_4$) do not have any particular pattern. We do not expect that general supervised learning can be used to distinguish them. Nevertheless, we can teach a machine to learn to \textit{recognize} positive sequences. Due to the need to classify time series, we use a typical method for modeling ordered data \cite{Hugheseaay2019, Rank2020, Tsai2020}, recurrent neural network (RNN) \cite{Hopfield1982, Hochreiter1997}. Specifically, we have designed a bidirectional long short-term memory (BLSTM) autoencoder (AE), whose goal is to reconstruct the input sequences as much as possible. We have trained this AE using positive sequences only and evaluated how well the AE reconstructs for an input sequence $\mathbf{x}$ using reconstruction error $\mathcal{L}(\mathbf{x})$ (see Methods Section \ref{ae_classifier},  Eq.\thinspace\ref{ae_loss}). After training, the autoencoder can reconstruct positive processes very well. However, when we input negative sequences into the AE, likely, it would not be able to reconstruct them well, leading to the reconstruction errors of these negative sequences greater than that of the positive sequences. Through the reconstruction error, we can determine whether a $\tilde{h}$ sequence is positive or negative. 
\begin{figure}[tbhp]
	\centering
	\includegraphics[width=\linewidth]{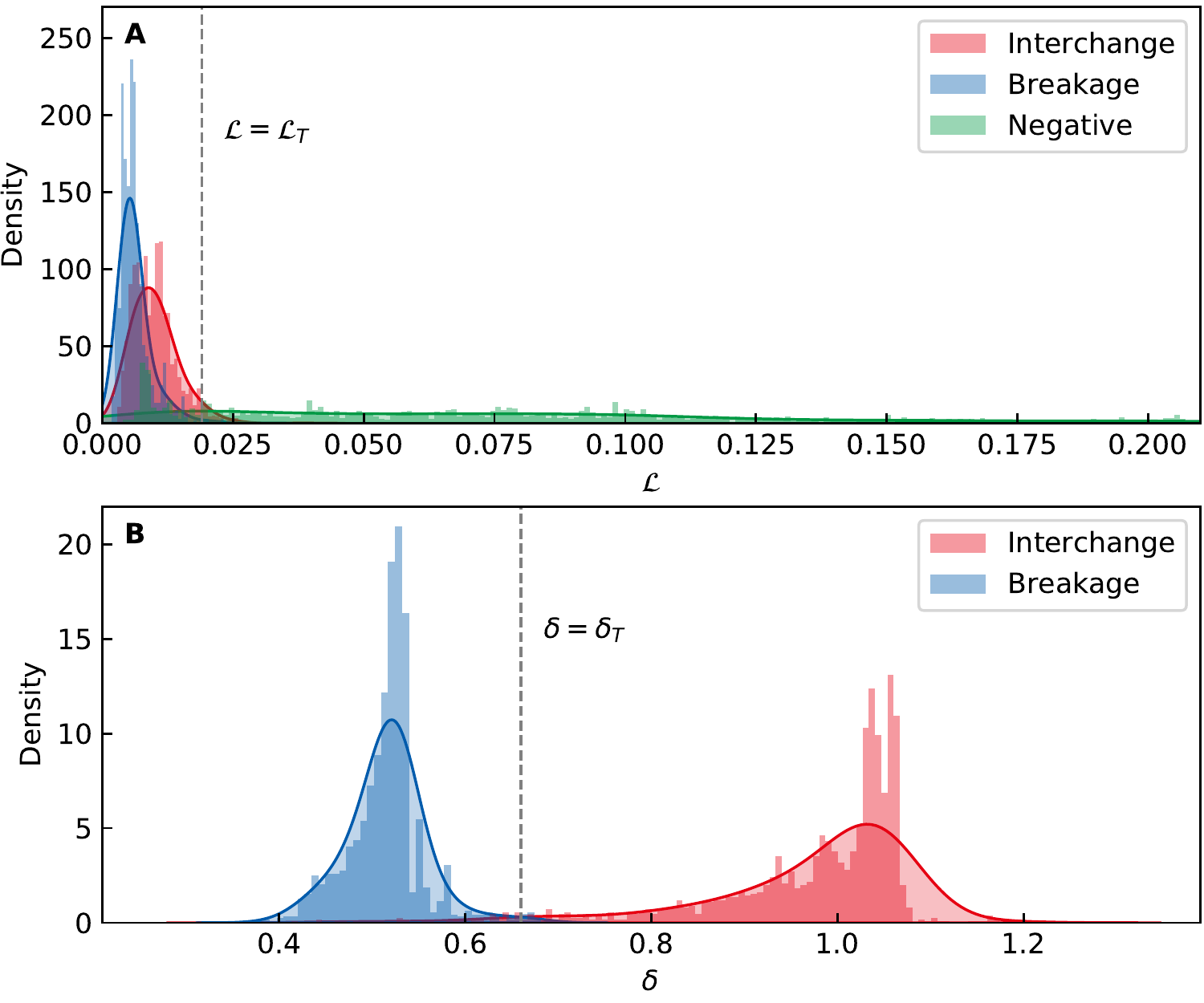}
	\caption{(A) Densities of reconstruction error $\mathcal{L}$ for interchange, breakage, and negative sequences. (i) BLSTM AE can reconstruct positive sequences well. Hence, the reconstruction errors for interchange and breakage sequences are relatively small, mainly less than $\mathcal{L}_\text{T}$. (ii) Since negative sequences are not used to train BLSTM AE, it is much more difficult for the autoencoder to reconstruct them. Therefore, the reconstruction errors  are relatively large, mainly greater than $\mathcal{L}_\text{T}$. (iii) Once $\mathcal{L}_\text{T}$ is  determined, we use it as the threshold to distinguish positive and negative sequences. (B) Densities of the range $\delta$ for interchange and breakage sequences. The two densities are significantly different from each other.}
	\label{fig:classifier_result}
\end{figure}
Finally, we use a \textit{final classifier} to distinguish sequences between interchange and breakage processes from positive sequences. We use the range of a positive sequence $\mathbf{x}$ to determine whether it is interchange or breakage, which is defined as $\delta (\mathbf{x}) = \max {\mathbf{x}} - \min  {\mathbf{x}}$. 

Figure \ref{fig:classifier_result} (A) shows the densities of the reconstruction errors for interchange, breakage, and negative sequences. Since BLSTM AE can reconstruct positive sequences well, the reconstruction errors of interchange and breakage sequences are small, most of which are smaller than the reconstruction error threshold $\mathcal{L}_\text{T}$ ($\mathcal{L} _\text{T}$ determination and corresponding accuracy analysis are described in SI Sec.\thinspace3, Fig.\thinspace$\text{S3}$).
Negative sequences are not used to train the autoencoder, so it is much more difficult to reconstruct them.  Hence, the reconstruction errors are relatively large, most of which are greater than $\mathcal{L}_\text{T}$. As long as we find a suitable reconstruction error threshold, we can get a classifier for positive and negative sequences. Figure \ref{fig:classifier_result} (B) shows the densities for the range of normalized interchange and breakage sequences. The two distributions are significantly different from each other. Therefore, the final classifier can distinguish interchange and breakage sequences very well via $\delta_\text{T} = 0.66$, as shown in the dashed line (see the classification process in SI Sec.\thinspace3, Fig.\thinspace$\text{S4}$-S5 ). Therefore, we have obtained an H-bond configuration change classifier based on an RNN autoencoder.

\subsection{Proportions of interchange at different temperatures}\label{ssec:proportions}
\begin{figure*}[thb]
	\centering
	\includegraphics[width=1\linewidth]{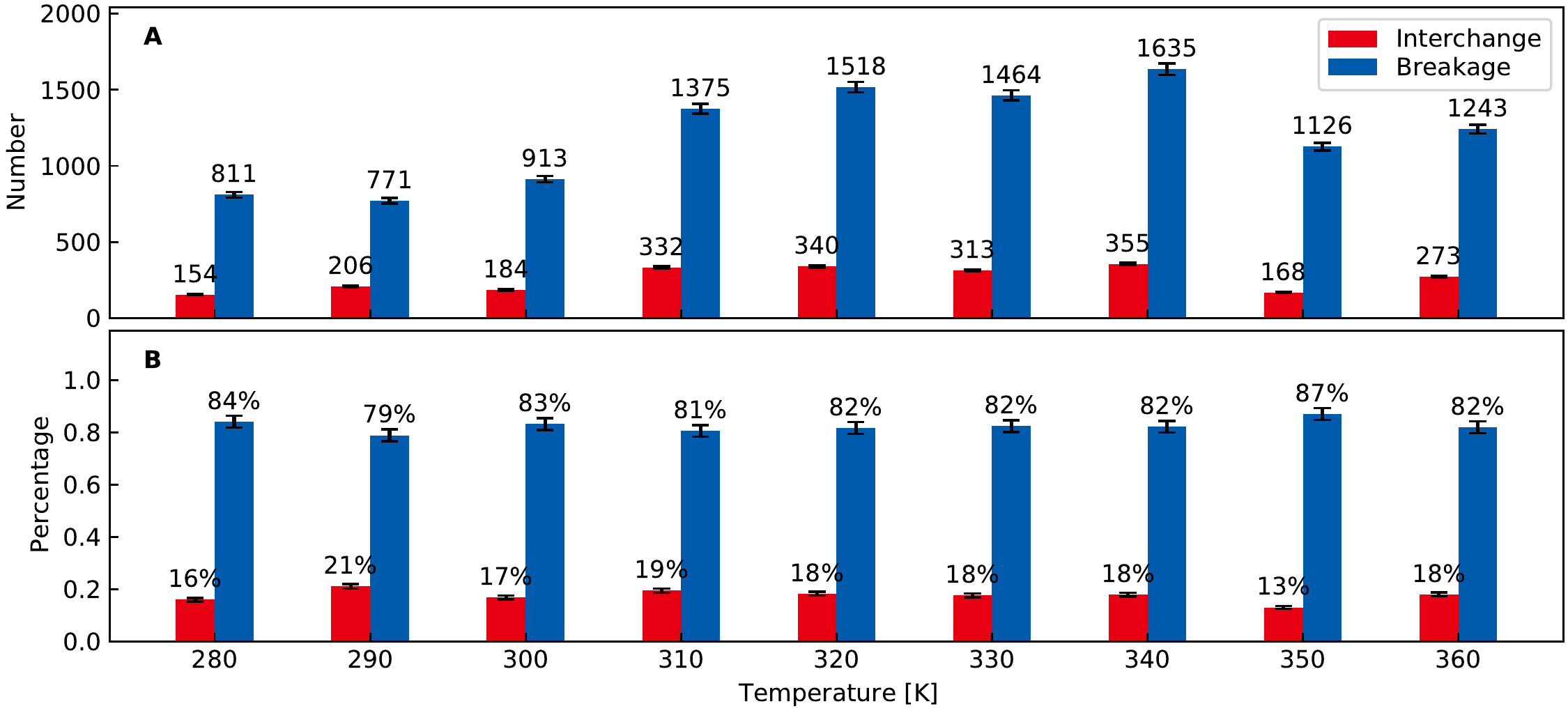}
	\caption{The number (A) and proportion (B) of interchange and breakage processes determined by the RNN-based classifier at different temperatures. (i) With the temperature increasing, the number of interchange and breakage processes increases first and then decreases on the whole. (ii) The relative ratio of interchange to breakage basically does not depend on temperature.
	}
	\label{fig:percentageK}
\end{figure*}
To explore the effect of temperature on the H-bond configuration change process, we have simulated nine bulk water systems containing $N=64$ water molecules. The temperature ranges from 280 to 360 K every 10 K. Using the RNN-based model, we classify  $\tilde{h}$ sequence, count the number of interchange and breakage sequences at each temperature. As shown in Fig.\thinspace\ref{fig:percentageK}, the number of interchange and breakage processes shows a "rising first, then decreasing" trend as the temperature increases. In other words, there is an overall upward trend from 280 to 330 K. However, as the temperature continues to rise, the number of detected interchange and breakage processes tends to decrease. As we use the method of width-fixed sliding window, the absolute number of interchanges and breakages would change along the step size of sliding window. These numbers would increase as we decrease the step size. Therefore, we focused on the trend of the detected processes over the temperatures. On the other hand, although the number of interchange and breakage processes vary at different temperatures, the relative ratios between the two are almost unchanged, which is still about 1:4 (see SI Sec.4 for the step size effect of the sliding window). This result indicates that the relative ratio is almost not dependent on temperature, and the interchange process is another important mechanism in bulk water besides the breakage process. Next, we will explain this trend of the number of interchange and breakage processes from the following two aspects: the number of H-bond per molecule and the change rate of the coarse-grained H-bond network configuration.

\subsection{The trend of interchange and breakage process number}\label{ssec:trend}
\begin{figure}[thb]
	\centering
	\includegraphics[width=\linewidth]{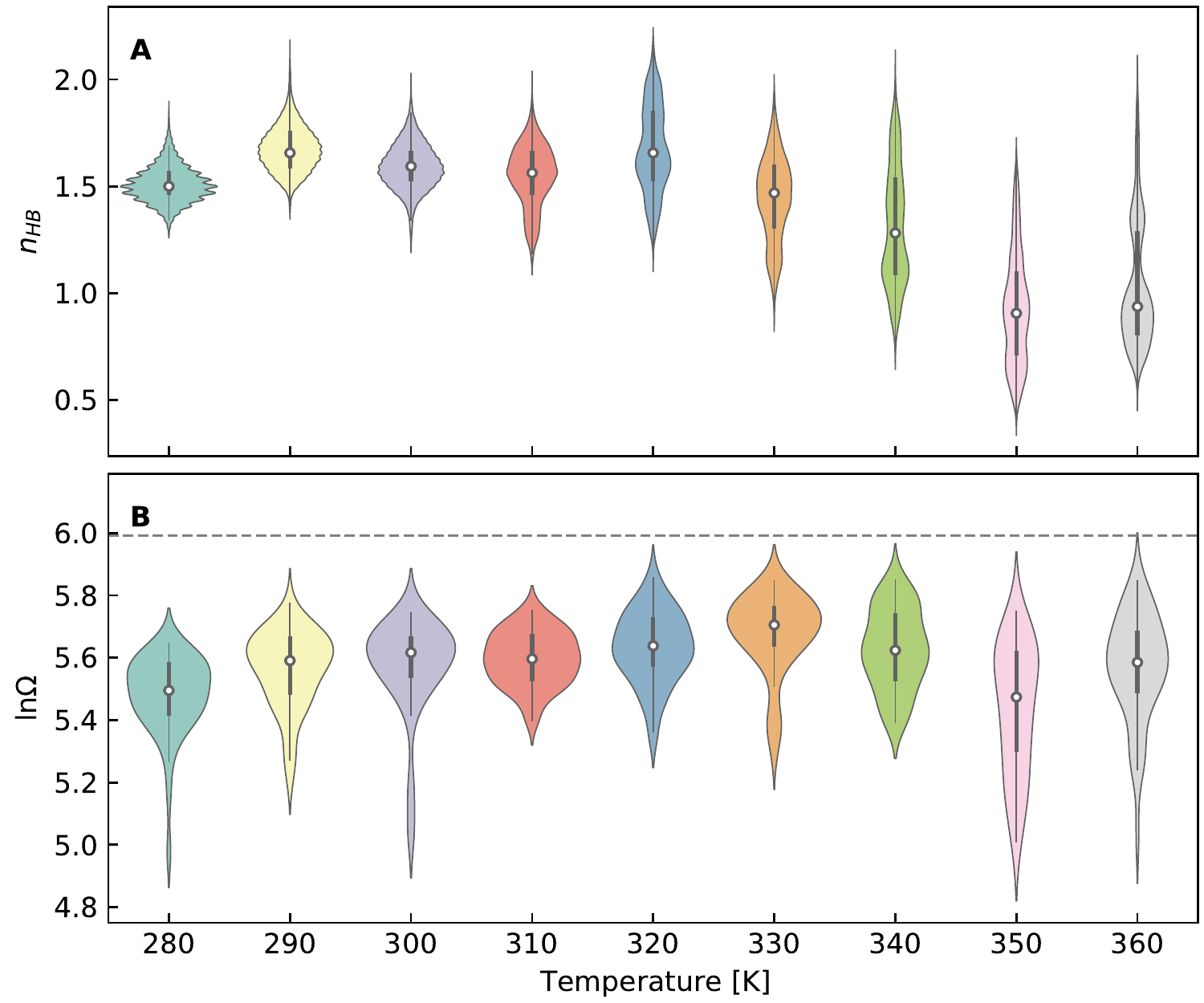}
	\caption{
		The temperature dependence of (A) The distributions of the number $n_{\text{HB}}$ of H-bonds per molecule. (B) The distributions of $\ln \Omega$ characterizing the rate of  H-bond breakage and reforming. The dashed line denotes the upper bound of $\ln \Omega$ in the unit time of 1 ps.}
	\label{fig:n_HB_omega}
\end{figure}

To understand the trend in Fig.\thinspace\ref{fig:percentageK} (A), we first calculate the number of H-bonds per molecule ($n_{\text{HB}}$) in the simulated system. At time $t$, $n_{\text{HB}}$ can be expressed as Eq.\thinspace\ref{eq:n_hb},
\begin{align}
	& n_{\text{HB}}(t) = \frac{2}{N}\sum_{i=1}^N\sum_{j>i}^{N} |\tilde{h}_{ij}(t)| \label{eq:n_hb}
\end{align}
where $N=64$ is the number of water molecules in bulk water systems, and $|\tilde{h}_{ij}(t)|$ is the absolute value of $\tilde{h}_{ij}(t)$, i.e., the H-bond direction is ignored. The factor 2 is derived from the fact that one H-bond in water is shared by two water molecules. For a certain trajectory at one temperature, by counting $n_{\text{HB}}$ at each time $t$, we get the distribution of $n_{\text{HB}}$ (one density plot in Fig.\thinspace\ref{fig:n_HB_omega} A).
Then we use an $L$-dimensional vector $\vect{\tilde h}(t)$  to represent the coarse-grained H-bond network  \textit{configuration} for the simulated bulk water system at time $t$ in Eq.\thinspace\ref{eq:vec_h},  
\begin{align}
	& \vect{ \tilde h}(t) =( \tilde h_{12}(t), \tilde h_{13}(t), \cdots, \tilde h_{ij}(t) , \cdots, \tilde h_{N-1,N}(t)) \label{eq:vec_h}
\end{align}
where $L=N(N-1)/2$  is the number of Q-bonds in the system. So in a unit time, we get a set $H$ of  $\vect{ \tilde h}(t)$ in Eq.\thinspace\ref{eq:set}, 
\begin{align}
	& H = \{ \vect{\tilde h} (t) \mid t = t_0 + k \Delta t,  k = 0, 1, \cdots, M \} \label{eq:set}
\end{align}
where $t_0$ represents the start time of the unit time window, $\Delta t$ is the time interval between two adjacent frames, and $M$ is the length of the unit time window. In a unit time $t_\text{w}=M \Delta t$, the number of graph configuration can be expressed as  $\Omega = |H|$, where $|H|$ is the size of the set $H$, i.e., the number of different $\vect{\tilde h}$ vectors  in this unit time. The number $\Omega$ of graph configuration per unit time characterizes the \textit{rate} of breakage and reforming of the H-bonds in bulk water. The theoretical upper bound of  $\Omega$ in $t_\text{w}$ is $M+1$; in this case, all $\vect{ \tilde h }$ vectors are different. By changing the start point $t_0$, we get the distribution of $\Omega$.

Figure \ref{fig:n_HB_omega} shows the temperature dependence of the distributions of  $n_{\text{HB}}$  and $\ln \Omega$. The width of a density plot indicates the probability of $n_{\text{HB}}$ or $\ln \Omega$ at the corresponding temperature. From the medians (white dots) of violin plots in Fig.\thinspace\ref{fig:n_HB_omega} (B),  we see $\Omega$ is relatively smaller at lower temperatures, indicating fewer changes of H-bond configuration in the unit time. This result explains why the number of interchange and breakage processes at lower temperatures in Fig.\thinspace\ref{fig:percentageK} (A) are smaller. Besides, the direct reason for the decrease in the number of interchange and breakage processes at higher temperatures is that thermal motions tend to break H-bonds (thus reducing $n_{\text{HB}}$). Therefore, the number of interchange and breakage processes in Fig.\thinspace\ref{fig:percentageK} (A) is determined by $n_{\text{HB}}$ and $\Omega$ together. 
	
\section{Methods}\label{methods}
\subsection{AIMD simulations}
AIMD simulations were carried out for bulk water of 64 water molecules within the canonical NVT ensemble using CP2K/QUICKSTEP (v7.1) \cite{CP2K}. The number $N$ of water molecules was 64 for all bulk water systems at different temperatures from 280 to 360 K. The length of the periodic cubic box was 12.4295 \AA. The discretized integration time step $\Delta t$ was set to 0.5 fs. The simulation time was 60 ps. The BLYP functional, which consists of Becke non-local exchange \cite{Becke1988} and Lee-Yang-Parr correlation \cite{LeeC1988}, was used; Interactions between the valence electrons and the ionic cores were described by GTH pseudopotentials \cite{Hartwigsen1998, Lippert1999}; Valence electrons were expanded in a basis set consisting of double-zeta Gaussian functions \cite{VandeVondele2007} and plane waves with a cutoff energy of 280 Ry \cite{CP2K}. The Nos$\acute{\text{e}}$-Hoover chain thermostat \cite{Martyna1992} was used to conserve temperature. DFT-D3 correction \cite{Grimme10} for the dispersion interaction was used to obtain a more accurate description of the vibrational properties. It is worth mentioning that the analysis method we proposed can be used on various simulation data, and the AIMD simulation used here is one of the options. The graph-based analysis method is independent of simulation data, so this method can be used to analyze more accurate simulation data in the future.

\subsection{Sequence collection and preprocessing}
The sequence length of $\tilde{h}$ was 200 corresponding to 8 ps simulation time. Positive sequences in which only one interchange or breakage process occurred were collected. Negative $\tilde{h}$ sequences used to evaluate the BLSTM AE classifier were also collected. BLSTM AE was trained by 6786 positive sequences, of which the interchange and breakage processes each accounted for half (754 positive sequences at each temperature). There were 18,931 negative sequences for evaluating the BLSTM AE classifier. The filtered sequence $\tilde{h}_{f}[n]$ was obtained by second-order Butterworth filter  implemented by Scipy \cite{Virtanen_scipy_2020}. In addition, if $\tilde{h}_{f}[0]-0.5 < 0.15$, indicating no H-bond at the beginning ($\text{ T}_1$). If the standard deviation $\sigma$ of $\tilde{h}_{f}[n]$ satisfy $\sigma < 0.1$, then we consider the H-bond configuration in the Q-bond has not changed ($\text{ T}_2$, $\text{ T}_3$).

\subsection{Bidirectional LSTM autoencoder classifier} \label{ae_classifier}
The encoder and the decoder of BLSTM AE can be expressed as two transformations,  $\phi: \mathcal{X} \rightarrow \mathcal{F}$ and $\psi: \mathcal{F} \rightarrow \mathcal{X}$, where $\mathcal{X}$ and $\mathcal{F}$ are the input space and the feature space, respectively. The dimension of $\mathcal{F}$ is smaller than that of $\mathcal {X}$, and the feature vector $\phi(\mathbf{x})$ is the compressed representation of input  $\mathbf{x}$. The input $\mathbf{x}$ of BLSTM AE is the normalized and filtered directed H-bond population operator sequence $\tilde{h}_f[n]$. The reconstruction error of BLSTM AE for a sequence $\mathbf{x} = \tilde{h}_f[n]$  is defined as
\begin{align}
	\mathcal{L}_{ \omega, \omega^\prime }\left( \mathbf{x} \right) = \left\|\mathbf{x}-\psi_{\omega^\prime}\left( \phi_{\omega}\left( \mathbf{x} \right)\right) \right\|^{2}\label{ae_loss}
\end{align}
where $\omega$, $\omega^{\prime}$ represent the parameters of the encoder and decoder respectively. The purpose of training is to obtain the optimal $\omega$, $\omega^{\prime}$,
\begin{align}
	&\omega^*, \omega^{\prime *} = \arg \min_{\omega, \omega^{\prime}} \frac{1}{m}\sum_{i=1}^{m} \mathcal{L}_{ \omega, \omega^\prime }(\mathbf{x}^{i})\label{ae_target}
\end{align}
where $\mathbf{x}^{i}$ represents the $i$-th sequence (SI, Fig. S1-S2). 

\section{Summary}\label{conclusions}

In summary, we have designed and trained a deep learning-based model to recognize different types of processes related to H-bonds. The priority of this model are its remarkable ability to classify different dynamic processes of water molecules and its wide range of applications to different kinds of simulation methods. The model can be transfered to other dynamic systems containning H-bonds with the form of $\text{O-H}\cdots \text{O}$. As a feasible example, combined with AIMD simulations, we have found that the relative ratio of interchange and breakage processes in bulk water is approximately 1:4, and this ratio hardly depends on temperature. 

Moreover, the key concepts used in this work are the dynamic graph and the newly defined directed H-bond population. This reasonable coarse-grained description of the H-bond network simplifies the analysis of H-bond dynamics dramatically. This work demonstrates that the semi-supervised RNN-based model has an outstanding capability of classifying the dynamic processes related to H-bonds in bulk water, which implies the great potential to extend our present scheme to distinguish more complex dynamic processes in other systems like the water-vapor interface and electrolyte solutions.

\begin{acknowledgments}
This research was supported by the National Natural Science Foundation of China (NSFC) (Grant No. 21973070) and the Graduate Scientific Research Foundation of Wenzhou University. The simulations were performed on the cluster in the College of Mathematics and Physics at Wenzhou University.
\end{acknowledgments}

\bibliography{ref}
	
\end{document}